\documentclass[prd,showpacs,preprintnumbers,nofootinbib,aps,9pt]{revtex4}
\usepackage{subeqnarray}
\usepackage{epsfig,amsmath,amssymb}
\usepackage{mathrsfs}
\usepackage[usenames,dvipsnames]{color}
\usepackage[pagebackref=true, colorlinks=true]{hyperref}
\definecolor{redish}{rgb}{0.7,0.2,0.0}  
\definecolor{bluish}{rgb}{0.2,0.5,0.8}
\hypersetup{linkcolor=redish,          
                  citecolor=blue,        
                  filecolor=magenta,      
                  urlcolor=bluish}          
\DeclareFontFamily{U}{rsfs}{}         
\DeclareFontShape{U}{rsfs}{m}{n}{<5> rsfs5 <6><7> rsfs7          %
  <8><9><10><10.95><12><14.4><17.28><20.74><24.88> rsfs10}{}     %
\DeclareMathAlphabet{\mathfs}{U}{rsfs}{m}{n}                     %
\newcommand{\mfs}[1]{\mathfs {#1}}                               %

\newcommand{\ba}{\nopagebreak[3]\begin{eqnarray}}
\newcommand{\ea}{\end{eqnarray}}
\newcommand{\bii}{\begin{itemize}}
\newcommand{\eii}{\end{itemize}}
\newcommand{\nn}{\nonumber}

\newcommand{\sO}{{\mfs O}}
\newcommand{\sA}{{\mfs A}}

\newcommand{\f}{\frac}

\def \d{\delta}

\def \l{\ell}
\def \g{\gamma}
\def \e{\epsilon}
\def \lp{\l_p}
\def \j{\sqrt{j(j+1)}}

\def \lm{\lambda}

\def \sj{s_j^{\star}}

\def \O{\Omega}
\def \o{\omega}

\def \({\left(}
\def \){\right)}
\def \[{\left[}
\def \]{\right]}
\begin{document}
\title{Black Hole Entropy from indistinguishable quantum geometric excitations}
\author{A. Majhi}%
\email{abhishek.majhi@gmail.com}
\affiliation{The Institute of Mathematical Sciences\\4th Cross St., CIT Campus, Taramani,\\Chennai, Tamil Nadu, India}

\begin{abstract}
In loop quantum gravity the quantum geometry of a black hole horizon consist of discrete non-perturbative quantum geometric excitations (or punctures) labeled by spins, which are responsible for the quantum area of the horizon. If these punctures are compared to a gas of particles, then the spins associated with the punctures can be viewed as single puncture area levels analogous to single particle energy levels. Consequently, if we {\it assume} these punctures to be {\it indistinguishable}, the microstate count for the horizon resembles that of Bose-Einstein counting formula for gas of particles. For the Bekenstein-Hawking area law(BHAL) to follow from the entropy calculation in the large area limit, the Barbero-Immirzi parameter($\g$) approximately takes a constant value. 
\end{abstract}
\maketitle
\section{Introduction}
One of the prime achievements of canonical quantum gravity, more specifically {\bf loop quantum gravity(LQG)}, is the provision of a description of microstates of equilibrium black hole horizon, modeled as quantum {\bf isolated horizon(IH)}\cite{qg1,qg2}, leading to an {\it ab initio} quantum statistical derivation of entropy from first principles. The quantum geometry of a cross-section of an IH is depicted as a topological 2-sphere with quantum degrees of freedom  at localized points called {\it punctures}. These punctures {\it are}  the quantum geometry and each puncture is associated with a quantum number $j$. 
If these punctures are considered to be distinguishable, then the microstate count for the quantum IH resembles that of Maxwell-Boltzmann(MB) counting for a gas of distinguishable particles\cite{gm,gp,am,ampm2}. In this case, if the microcanonical entropy has to be given by {\bf Bekenstein-Hawking area law(BHAL)}(i.e. one-fourth of the area of the horizon divided by Planck length squared), then the {\bf Barbero-Immirzi parameter$(\g)$} needs to take a certain fixed value.

However, recently there has been a trend of doing black hole thermodynamics in the quantum IH framework considering the punctures to be {\it indistinguishable}\cite{gnp,achour,pithis,mhan}. While in some cases Gibbs's approximation has been used in the MB counting to implement indistinguishability\cite{mhan}, the others use both Bose-Einstein(BE) and Fermi-Dirac(FD) statistics\cite{gnp,achour} by treating the punctures {\it literally} as quantum mechanical particles with spins and thus categorizing the punctures with integral and half-integral spins into two different species which follow BE and FD statistics respectively. Also, there is an instance where anyonic statistics has been used\cite{pithis}.

In this article, we shall revisit this issue of indistinguishability of punctures of quantum IH. We {\it assume} that the punctures are {\it indistinguishable}. Consequently, the microstate count now {\it resembles} that of BE statistics applied to a gas of particles.  We explain that the resemblance between a quantum IH and a gas of particles follows if we view the quantum number $j$ as denoting the area levels of an individual puncture, rather than `spin', similar to the energy levels of an individual particle in a gas. Automatically, this provides an  explanation of how our work differs from other instances in literature where BE statistics has been discussed in relation to quantum IH by considering the quantum number $j$ as `spins' associated with the punctures (e.g. see \cite{gnp}). Also, we explain why we think it is more reasonable to consider the quantum number $j$ as area levels of the individual punctures rather than `spin' analog of a particle by pointing out some fundamental difference between the physics associated with a quantum IH and a gas of particles in spite of the structural similarity in their statistical mechanical framework.
Then, we calculate the microcanonical entropy of an IH with a given classical area $A\gg\sO(\lp^2)$ (where $\lp$ is the Planck length) and find the restrictions to be imposed on the BI parameter for the BHAL to follow. We find graphically that the BI parameter has  an area dependence, but approaches a constant value in the large area limit. Finally we end with some concluding remarks.


\section{Microstate Count}

An individual puncture of a quantum IH labeled with quantum number  $j$ contributes a quantum of area $a_j$ to the quantum IH. If $s_j$ denotes the number of punctures with label $j$, then, for a set $\{s_j\}$, the total quantum area is given by  $A_{quant}=\sum_js_ja_j$. Also, $g_j$ is the degeneracy associated with a puncture labeled by quantum number $j$. 

On the other hand, let us consider a gas of particles with total energy $\sum_i n_i \e_i$, where $i\equiv$ {\it single particle energy level} (SPEL) , $\e_i\equiv$ energy of a particle in the $i$-th level, $n_i\equiv$ number of particles in the $i$-th level. $\omega_i$ be the degeneracy associated with the $i$-th level.

There is a manifest structural similarity between these two systems if we consider the following correspondence $(\cal G)$ : 
\ba 
j\longleftrightarrow i~,~s_j\longleftrightarrow n_i~,~ a_j\longleftrightarrow \e_i~,~ g_j\longleftrightarrow \omega_i\nn
\ea
Thus, $j$ can be called as {\it single puncture area level} (SPAL) for a  quantum IH.  

As the underlying quantum theory of the gas of particles provide the details of $\e_i, \o_i$, etc. , so does LQG for a quantum IH viz. $a_j=8\pi\g\lp^2\j$,  $g_j=(2j+1)$ and $j$ can take values in the range $1/2,1,3/2,\cdots,A/8\pi\g\lp^2$ \cite{qg1,qg2}. 

Now, for a quantum IH, $s_j$ can vary arbitrarily without any restrictions  and the punctures are indistinguishable by assumption.
Hence, considering the correspondence $\cal G$, the system has a structural similarity with a gas of indistinguishable particles where $n_i$ can vary arbitrarily without any restrictions. The microstate count for a configuration $\{n_i\}$ is well known\cite{pathria}. By the correspondence $\cal G$, we can simply get the number of microstates  corresponding to a configuration  $\{s_j\}$ of a quantum IH , which is given by 
\ba
\O[\{s_j\}]=\prod_j\f{(s_j+g_j-1)!}{s_j!~(g_j-1)!} \label{count}
\ea

[{\it Digression :}  If we consider the punctures to be distinguishable and since, any number of punctures can have any value of $j$, the microstate count for a set  $\{s_j\}$ of punctures is given by $ \Omega[\{s_j\}]=(\sum_ks_k)\prod_j(g_j^{s_j}/s_j!)$\cite{gm,am,ampm2}, which  resembles  Maxwell-Boltzmann(MB) counting for a gas of particles considering the correspondence $\cal G$ \cite{pathria}.]

The counting details are available in standard textbooks of statistical mechanics (e.g. see \cite{pathria}) and need not be discussed here unnecessarily. However, we shall cross-check the above formula with a simple example for a clarification. Let us consider that there are two punctures with $j=1/2$ and three punctures with $j=1$. So the corresponding degeneracies are -1/2,1/2 and -1,0,1 respectively. Since the punctures are indistinguishable, the distinct microstates that can be constructed out of the two $j=1/2$ punctures are $(1/2,1/2), (1/2,-1/2)$ and $(-1/2,-1/2)$. Similarly, for the three $j=1$ punctures the distinct microstates are $(1,1,1), (1,1,0),(1,0,0),(0,0,0),(0,0,-1),(0,-1,-1),(-1,-1,-1),(-1,-1,1),(-1,1,1)$ and $(-1,0,1)$. Hence, there are $3\times 10 = 30$ distinct microstates of the system of five indistinguishable punctures. Now, if we put $s_{1/2}=2, g_{1/2}=2$ and $s_{1}=3,g_1=3$, then the eq.({\ref{count}}) yields $30$. So, the formula given by eq.(\ref{count}) stands justified.  

This completes our effort to explain that, in what precise sense the punctures of a quantum IH obey BE statistics under the assumption of indistinguishability. However, we need to further clarify certain other issues regarding the analogy between a quantum IH and a gas of particles in order to differentiate our view point from other instances in literature which discuss BE statistics in relation to quantum IH.

From the correspondence $\cal G$ between a quantum IH and a gas of particles, it is manifest that there is no corresponding analog of  spin of a particle for a puncture of quantum IH. The quantum number $j$ is called as `spin' of a puncture in literature because it originates from the SU(2) spin representation carried by the edge of the bulk spin network graph that ends on the corresponding puncture\cite{qg1,qg2}. But we have already explained that $j$ actually represents the SPAL of a quantum IH when it comes to microstate counting. 

The spins of particles do not explicitly appear in the microstate counting. It only implicitly affects the counting by imposing restriction  on $n_i$ through spin-statistics connection. On the other hand, any number of punctures can be associated with integral or half-integral $j$-s in case of quantum IH. Alternatively, it can be said that there are no other independent quantum number (analog of particle spin) associated with the punctures of a quantum IH which imposes any restriction on $s_j$ implicitly. 



In spite of all the above facts one can still wish to treat the quantum number $j$ as `spins' and punctures as particles. In that case, one needs to show how this treatment invokes the spin-statistics connection and affects the microstate count of a quantum IH. The main obstacle in taking this view point is the deep mismatch between the fundamental natures of punctures of quantum IH and quantum mechanical particles. Particles are perturbative quantum excitations of matter fields propagating on smooth background geometry\cite{qft}, whereas the punctures are non-perturbative background independent quantum geometric excitations of the IH\cite{qg1,qg2}.  The spin-statistics connection heavily relies on local Lorentz invariance related to the background spacetime. But, punctures of quantum IH being themselves the representatives of the quantum geometry, the notion of spin-statistics connection is hitherto unknown in this scenario.


\section{Entropy}
Now, we shall calculate the entropy of an IH with classical area $A\gg\sO(\lp^2)$. The physical quantum states of the quantum IH which are of interest in this particular calculation are such that the quantum area corresponding to each state is within $\pm\sO(\lp^2)$ window of the classical area $A$ i.e. in mathematical language $A_{quant}[\{s_j\}]=A\pm\sO(\lp^2)$. As we are working with $A\gg\sO(\lp^2)$, we can neglect $\pm\sO(\lp^2)$ for all practical purposes. Thus, the relevant  configurations that will contribute to the entropy must obey the constraint 
\ba
C:~~8\pi\g\sum_{j}s_j\j=\sA
\ea
where $\sA=A/\lp^2$ is the dimensionless area of the quantum IH.
So the entropy of the IH, using Boltzmann entropy formula and setting Boltzmann constant to unity, is then given by
\ba
S=\log \[\sum_{\{s_j\}} \O[\{s_j\}]\]
\ea
where the argument of the logarithm represents the total number of microstates arising from all possible configurations, the sum being over all possible configurations  constrained by $C$. {\it Assuming} that the basic postulates of equilibrium statistical mechanics are valid in present scenario of quantum IH, there is one most probable configuration whose corresponding number of microstates is overwhelmingly large compared to any other configuration\cite{pathria} such that the entropy of the IH can be approximately given by the entropy of the most probable configuration alone i.e.  
\ba
S&=&\log \[\sum_{\{s_j\}} \O[\{s_j\}]\]\nn\\
&=&\log \O[\{\sj\}] + \text{contributions from the sub-dominant configurations}\nn\\
&\simeq&\log \O[\{\sj\}]
\ea
The most probable configuration can be obtained by extremizing the entropy corresponding to a configuration subject to the area constraint $C$ and hence it is the solution of the following equation :
\ba
\d \log \O[\{s_j\}]-\lm \d \sA=0
\ea
where $\d$ represents arbitrary variation with respect to the variable $s_j$. The distribution function for the most probable configuration (or the most probable distribution)  comes out to be
\ba
\sj= \f{(g_j-1)}{e^{8\pi\lm\g \j}-1}= \f{2j}{e^{8\pi\lm\g \j}-1}\label{sjstar}
\ea
Considering $s_j\gg1$ and consequently applying Stirling's approximation, the entropy is calculated to be 
\ba
S &\simeq & \log \O[\{\sj\}]\nn\\
&=&\lm \sA-\sum_{j=1/2}^{\sA/8\pi\g}2j\log\(1-e^{-8\pi\lm\g \j}\)+\sum_{j=1/2}^{\sA/8\pi\g}\psi_j \label{ent}
\ea
where the upper limit on $j$ is written in terms of $\sA$ i.e. $j_{max}=A/8\pi\g\lp^2=\sA/8\pi\g$ and here
\ba \psi_j&=&\log(g_j-1)!-(g_j-1)\log(g_j-1)+(g_j-1)\nn\\
&=&\log(2j)!-2j\log(2j)+2j
\ea
The most probable distribution satisfies the area constraint which leads to the following equation
\ba
\sA&=&8\pi\g\sum_{j=1/2}^{\sA/8\pi\g}\f{\j(g_j-1)}{e^{8\pi\lm\g \j}-1}=16\pi\g\sum_{j=1/2}^{\sA/8\pi\g}\f{j\j}{e^{8\pi\lm\g \j}-1}\label{area}
\ea
Besides this, summing over $j$ gives the total number of punctures for the most probable distribution i.e. 
\ba
N_0&:=\sum_j\sj=&\sum_{j=1/2}^{\sA/8\pi\g}\f{(g_j-1)}{e^{8\pi\lm\g \j}-1}=\sum_{j=1/2}^{\sA/8\pi\g}\f{2j}{e^{8\pi\lm\g \j}-1}
\ea
It may be noted that the allowed values of $\lm$ and $\g$ must be such that $N_0\gg 1$. This condition will be automatically satisfied if, at least, $\sj|_{max}\gg 1$ is assured for the allowed values of $\lm$ and $\g$. By $\sj|_{max}$ we have meant the peak values of the curve $\sj$ vs $j$ for given values of $\lm$ and $\g$. As we proceed, we shall see that there exist such values of $\lm$ and $\g$ for which the condition is satisfied.

\section{The BHAL and the Barbero-Immirzi parameter}
\begin{figure}[hbt]
\begin{center}
\includegraphics[scale=0.5]{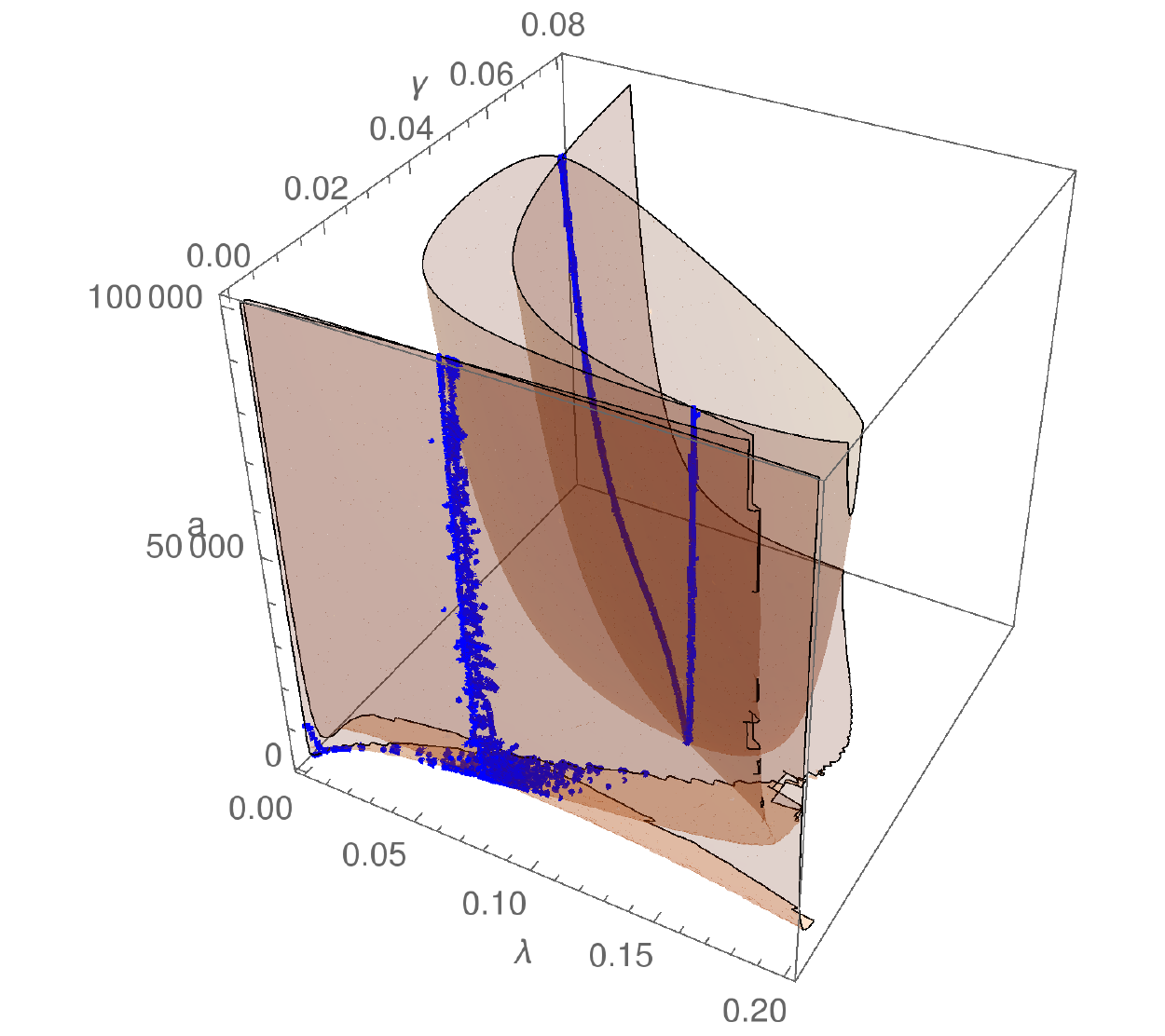}
\caption{\label{3dfig}The blue curves are the solutions of the eq.(\ref{area}) and eq.(\ref{alaw}). $\sA$ is symbolized as `$a$' in the 3d plot. There are two branches of solutions, of which the U-shaped curve is excluded on physical grounds. The other curve is the only feasible solution which has a broadening due to graphical impreciseness of the plot. (See {\it Appendix}.)}
\end{center}
\end{figure}

The next step is to choose $\g$ in such a way that eq.(\ref{ent}) yields the BHAL. Hence, we demand that the entropy should be given by the BHAL i.e. 
\ba
S=\sA/4\label{arealaw}
\ea
and then explore the consequences.  Now, demanding eq.(\ref{arealaw}) we obtain the following equation 
\ba
\lm \sA-\sum_{j=1/2}^{\sA/8\pi\g}2j\log\(1-e^{-8\pi\lm\g \j}\)+\sum_{j=1/2}^{\sA/8\pi\g}\psi_j= \sA/4\label{alaw}
\ea
Our motto is to show that there exists certain value(s) of $\g$ for which the BHAL can be derived from the indistinguishable punctures, which is equivalent to finding the solutions of eq.(\ref{area}) and eq.(\ref{alaw}). This can be accomplished by finding if there is any  intersection of the 2-surfaces described by eq.(\ref{area}) and eq.(\ref{alaw})in the 3-space spanned by $\sA, \lm$ and $\g$. Such a plot in Mathematica, shown in Fig.(\ref{3dfig}), reveals that there indeed exist values of $\g$, which furthermore fulfill certain relevant criteria as follow :
\begin{itemize}
\item $\g$ should lie in the range $0<\g\leq\sA/4\pi$ because $j_{max}=\sA/8\pi\g\geq 1/2$ and $\g$ appears in the area spectrum as a multiplicative factor, hence needs to be positive definite.
\item The allowed values of $\g$ are such that we have $\sA\gg\sO(1)$ and $\lm>0$ for $\sj$ to be positive definite for all $j$ (see eq.(\ref{sjstar})).
\end{itemize}  
On physical grounds we shall exclude the U-shaped curve from consideration because it gives rise to multiple values of $\g$ for a given $\sA$ which imply the following unphysical result : for a given $\sA$ we will have two different most probable spin distributions. 
\begin{figure}[hbt]
\begin{center}
\includegraphics[scale=0.5]{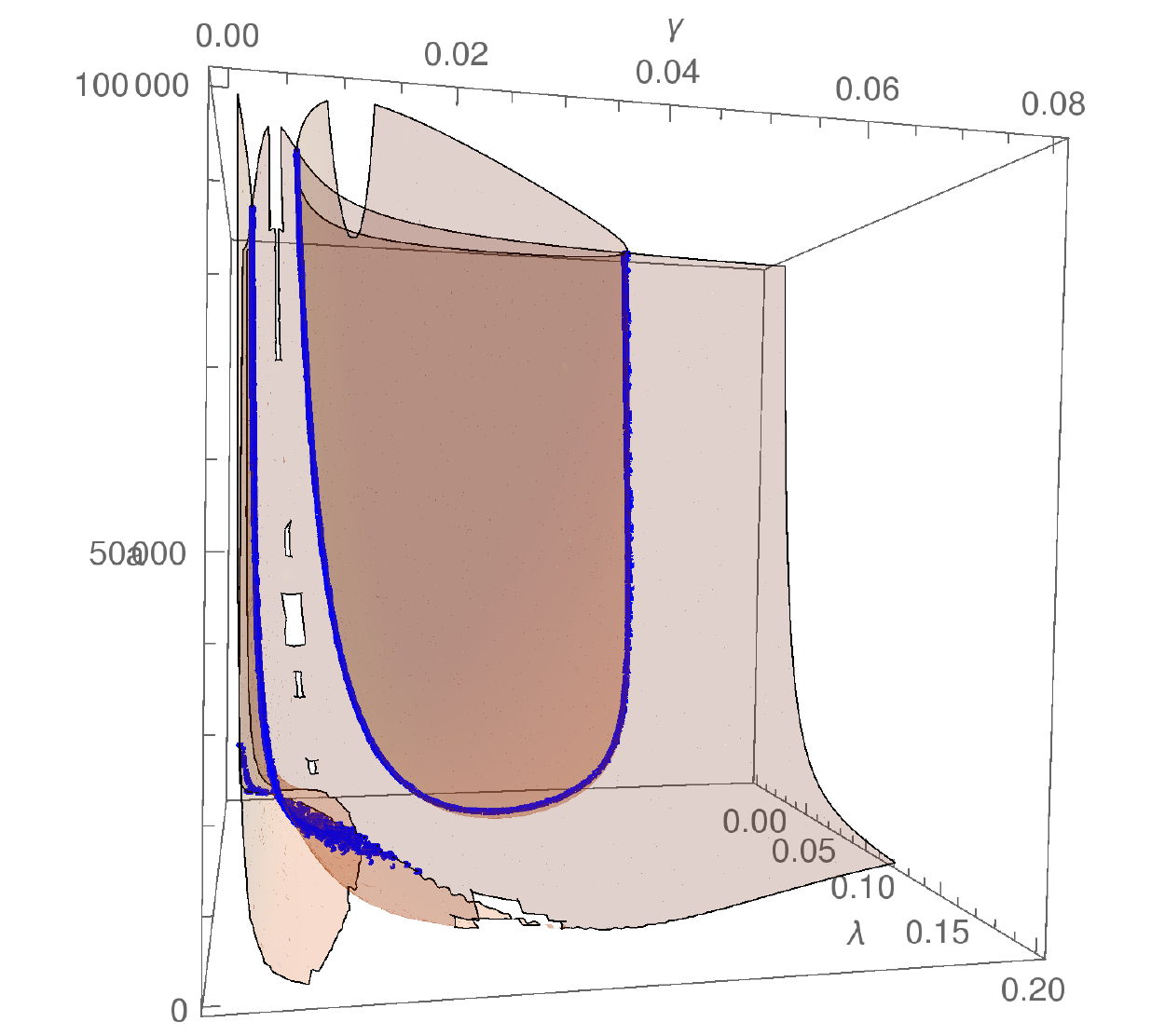}
\caption{\label{3dfig2}The view of the plot in FIG.(\ref{3dfig}) along the $\lm$-axis reveals the variation of $\g$ with $\sA$ (symbolized as $a$ in the figure). As $\sA$ increases, the value of $\g$ asymptotes to a fixed value $\g_0$(say) of the order of $10^{-3}$. (See {\it Appendix}.)} 
\end{center}
\end{figure}
So we are left with the other curve which will give us the value(s) of $\g$ for which the BHAL is valid. From FIG.(\ref{3dfig2}), which gives us the view of 3d plot along the $\lm$ axis, we see that with increasing area $(\sA)$ the curve straightens very rapidly and asymptotes to some constant value $\g_0$ (say). Besides that, from FIG.(\ref{3dfig}) it was already evident that $\lm$ becomes $\sA$ independent with increasing $\sA$. Hence, for $\sA\gg\sO(1)$, we have practically $\g=\g_0$ and $\lm=\lm_0$ for the BHAL to hold. In view of this, we can conclude that the BHAL follows from the LQG description of black hole horizon, with the same consequences whether we consider the punctures to be distinguishable or not. The only difference is the value of $\g_0$ which is of the order of $10^{-3}$ here, but of the order of $10^{-1}$ in case of distinguishable punctures. It may be further noted that the order of magnitude of $\g$ and $\lm$ along the curve ensure  that at least $\sj |_{max}\gg 1$ so that the Stirling's approximation is valid. One can check that to remain assured.

\section{Conclusion}
What we have done here is the entropy calculation of the IH by assuming that the punctures of quantum IH are indistinguishable and we do not go into the debate whether this assumption is justified within this quantum IH framework. As far as the physical result is concerned i.e. the BHAL resulting from an {\it ab initio} statistical mechanical calculation from the quantum geometry of black hole horizon as laid down by LQG, there is not much difference whether we consider the punctures to be distinguishable or not. Given that the calculations are done for large area black holes, the BHAL is obtained for a constant value of $\g$, except the fact that order of magnitude of  $\g$ is different in case of distinguishable and indistinguishable punctures.

   However, there is a very important issue which we are unable to address here in the context of the assumption of indistinguishability. The states of the quantum IH are actually given by that of the quantum SU(2) Chern-Simons theory coupled to the punctures\cite{km98}. Now, in the case of distinguishable punctures, it can be shown that the MB counting results from a zeroth order approximation of the complete formula for the state counting\cite{ampm2}. The distinguishability is inherited by the fact that we consider a sum over all values of spins on the microstate count for a given set of spins $j_1,\cdots, j_N$ and arrive at the MB counting for a given spin configuration by the application of multinomial expansion\cite{ampm2}. The next order correction leads to the logarithmic correction\cite{sigma,ampm2}. Now, when the punctures are indistinguishable, we can not carry out a naive sum over all values of spins for a given puncture data as this will count the microstates which are meant to be indistinguishable. Thus, under the assumption of indistinguishability of punctures, it will be an interesting problem to look upon how the BE counting formula, as discussed here, comes out as an approximation of the exact scenario of quantum Chern-Simons theory. Then it will also be clear what the next order correction is. Will it be the same known logarithmic correction $-3/2 \log \sA$ \cite{km00} ? We hope to investigate this problem in future.

\vspace{0.2cm}
{\bf Appendix :}
Here we have presented some magnified versions of the 3d plot as follows.
\begin{figure}[hbt]
\begin{center}
\includegraphics[scale=0.7]{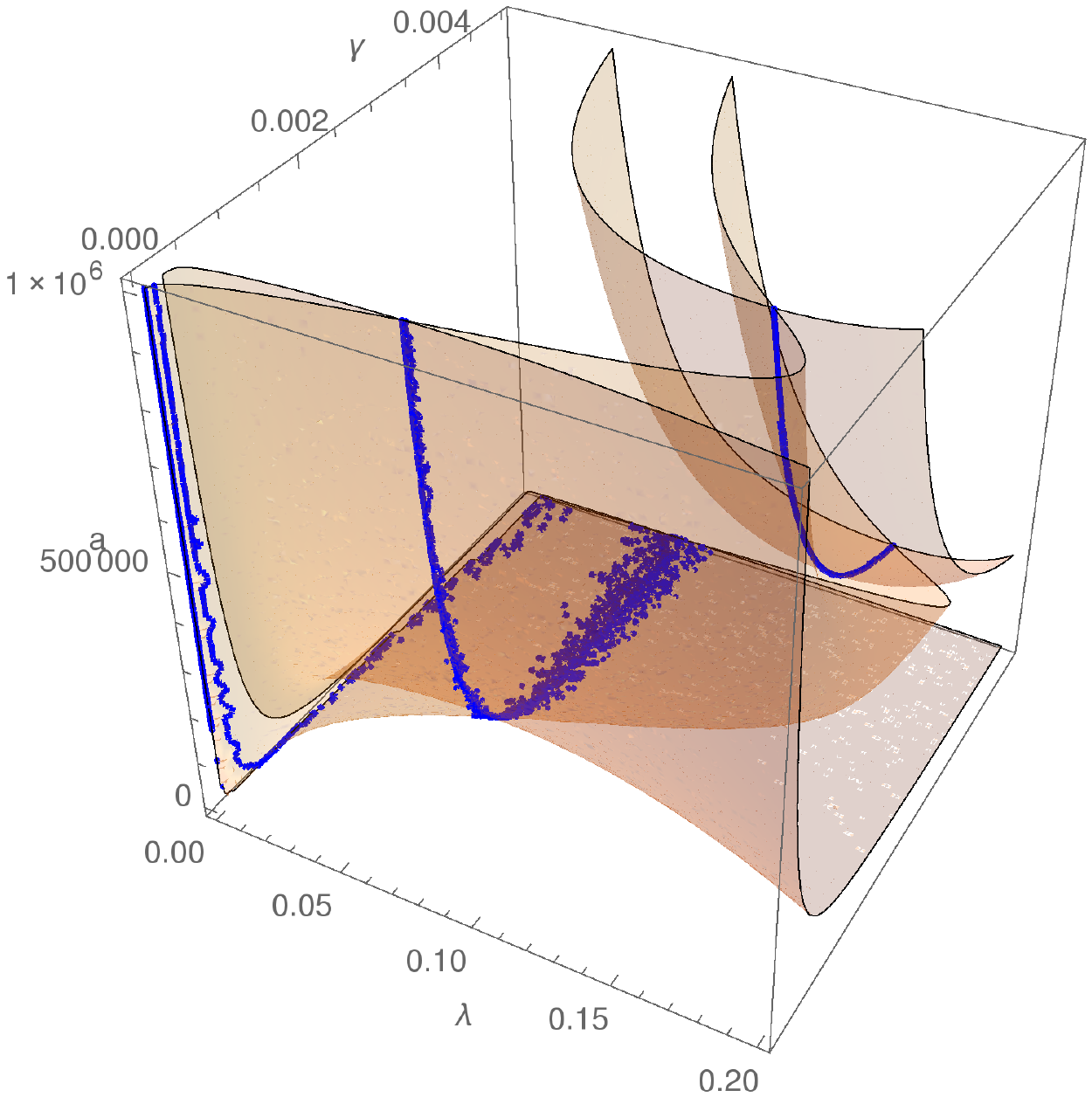}
\caption{\label{3dfig3} This is a magnified view of the plot in FIG.(\ref{3dfig}) to focus on the curve which gives the solution (along midway of the $\lm$-axis). }

\includegraphics[scale=0.7]{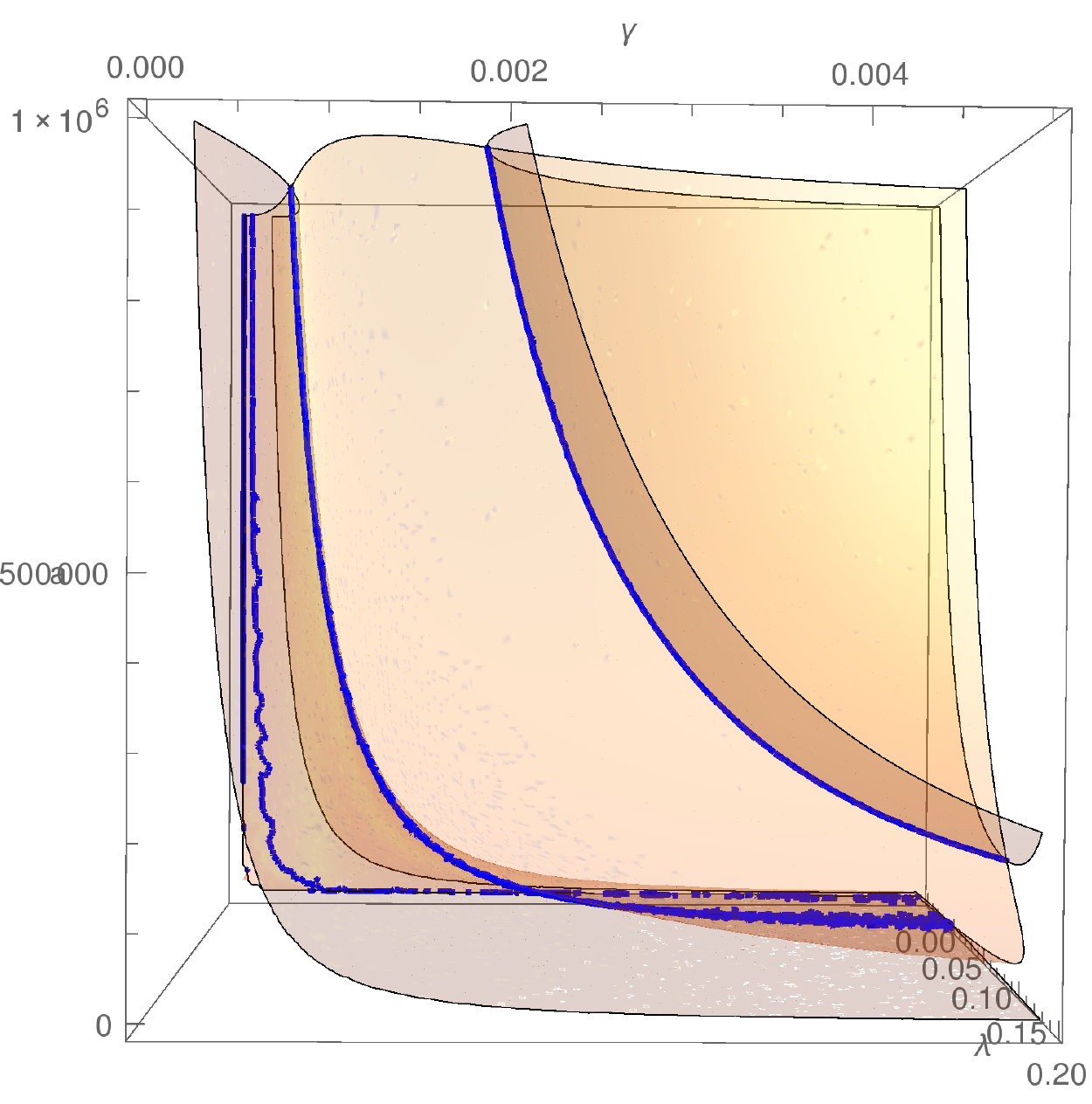}
\caption{\label{3dfig4} This is a magnified view of the plot in FIG.(\ref{3dfig2}) to show the view along $\lm$-axis more closely.}
\end{center}
\end{figure}

\vspace{0.5cm}
{\bf Acknowledgment :} The author thanks Romesh Kaul for discussions regarding a few issues of this work. This work is funded by The Department of Atomic Energy, India.

\end{document}